# A GOES Imager-Derived Microburst Product

Kenneth L. Pryor
Center for Satellite Applications and Research (NOAA/NESDIS)
Camp Springs, MD

## 1. Introduction

A suite of products has been developed and evaluated to assess meteorological hazards to aircraft in flight derived from the current generation of Geostationary Operational Environmental Satellite (GOES). The existing suite of products includes derived images to address major aviation hazards including fog, aircraft icing, and volcanic ash.  The products, derived from the GOES imager, utilize algorithms that employ temperature differencing techniques to highlight regions of elevated risk to aircraft. A new multispectral GOES imager product has been developed to assess downburst potential over the western United States.  Wakimoto (1985), based on his study of microbursts that occurred during the Joint Airport Weather Studies (JAWS) project, noted favorable environmental conditions over the western United States: (1) intense solar heating of the surface and a resulting superadiabatic surface layer; (2) a deep, dry-adiabatic convective boundary layer (Sorbjan 1989) that extends upward to near the 500mb level; (3) a well-mixed moisture profile with a large relative humidity gradient between the mid-troposphere and the surface.  Peak downdraft speeds associated with microbursts over the western U.S. result from negative buoyancy due to evaporation of precipitation during descent below cloud base.

Transmittance weighting functions specify the relative contribution each atmospheric layer makes to the radiation emitted to space and thereby determine regions of the atmosphere which are sensed from space at a particular wavelength. Temperature and moisture profiling for the purpose of inferring the presence of a favorable environment for microbursts would be accomplished with a group of spectral bands selected to detect radiation emitted from layers of interest in the atmosphere.  GOES-11 imager channel characteristics are available online: http://cimss.ssec.wisc.edu/goes/goesmain.html#imgrinfo .  As illustrated in Figure 1, transmittance weighting functions over the northwestern United States in the summer would dictate that band 3 is most sensitive to water vapor present between the 200 and 500 mb levels.  Band 5 is more sensitive to moisture in the boundary layer, and thus, when significant moisture is present, brightness temperatures observed by band 5 would be slightly lower than that observed in band 4.

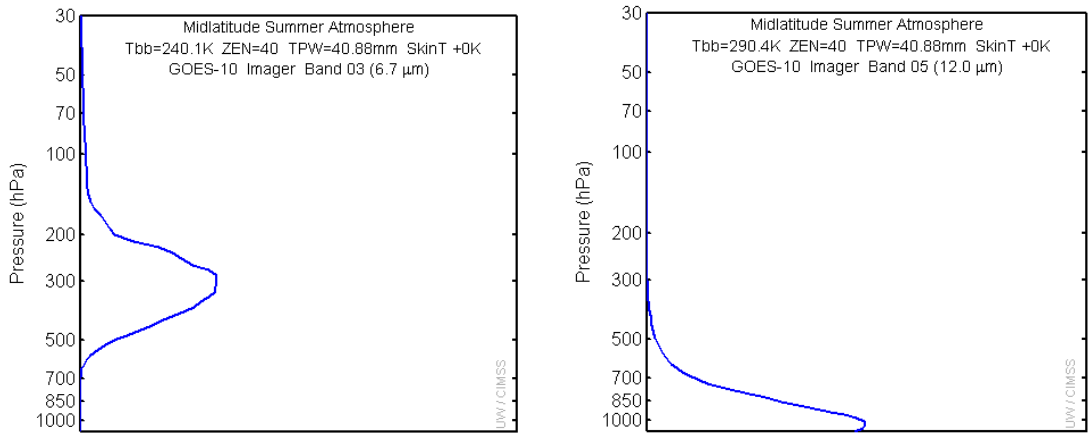

Figure 1. Transmittance weighting functions for GOES-10/11 bands 3 and 5.

Accordingly, a new GOES-West (GOES-11) imager microburst algorithm employs brightness temperature differences (BTD) between band 3 (upper level water vapor, 6.7μm), band 4 (longwave infrared window, 10.7μm), and split window band 5 (12μm). Band 3 is intended to indicate mid to upper-level moisture content and advection while band 5 indicates low-level moisture content. It follows that large BTDs between bands 3 and 5 imply a large relative humidity gradient between the mid-troposphere and the surface, a condition favorable for strong convective downdraft generation due to evaporational cooling of precipitation in the deep sub-cloud layer. In addition, small BTDs between bands 4 and 5 indicate a relatively dry surface layer with solar heating in progress. Thus the GOES imager microburst risk (MBR) product is based on the following algorithm in which the output brightness temperature (B) is proportional to microburst potential:

$$\text{MBR (B)} = \{T_5 - T_3\} - \{T_4 - T_5\} \tag{1}$$

Where the parameter $T_n$ represents the brightness temperature observed in a particular imager band. The relationship between BTDs and microburst risk in the product image is based on the following assumptions: (1) A deep, well-mixed convective boundary layer exists in the region of interest; (2) moisture for convective storm development is based in the mid-troposphere and is advected over the region of interest.

This paper will outline the development of the GOES-West imager microburst product and present case studies that feature example images, outline potential operational use and assess performance of the algorithm.

## 2. Methodology

The objective of this validation effort was to qualitatively and quantitatively assess the performance of the GOES imager-derived microburst product by employing classical statistical analysis of real-time data. Accordingly, this effort entailed a study of downburst events over the Eastern Snake River Plain (ESRP) of southeastern Idaho during the summer of 2007 that was executed in a manner that emulates historic field projects such as the 1982 Joint Airport Weather Studies (JAWS) (Wakimoto 1985). GOES-11 image data was collected for pre-convective environments associated with eight microburst events that occurred within the Idaho National Laboratory (INL) mesonet domain during July and August 2007.  Clawson et al. (1989) provides a detailed climatology and physiographic description of the INL as well as a description of the associated meteorological observation network.  Wakimoto (1985) and Atkins and Wakimoto (1991) discussed the effectiveness of using mesonet surface observations and radar reflectivity data in the verification of the occurrence of downbursts. Well-defined peaks in wind speed as well as significant temperature decreases (Wakimoto 1985; Atkins and Wakimoto 1991) were effective indicators of high-reflectivity downburst occurrence.  Product images were generated in Man computer Interactive Data Access System (McIDAS) by a program that reads and processes GOES imager data, calculates brightness temperature differences, and overlays risk values on GOES imagery. The image data consisted of derived brightness temperatures from infrared bands 3, 4, and 5, obtained from the Comprehensive Large Array-data Stewardship System (CLASS, http://www.class.ncdc.noaa.gov/).  In addition, visible imagery was generated to compare derived product images based on infrared bands to the location and spatial coverage of cloud fields prior to microburst events.  It was found that derived product imagery generated two to three hours prior to microburst events provided an optimal characterization of the pre-convective thermodynamic environment over the INL.

      Downburst wind gusts, as recorded by National Oceanic and Atmospheric Administration (NOAA) mesonet observation stations, were measured at a height of 15 meters (50 feet) above ground level. Archived NOAA mesonet observations are available via the NOAA INL Weather Center website (http://niwc.noaa.inel.gov).  In order to assess the predictive value of the GOES imager microburst product, the images used in validation were obtained for valid times two to three hours prior to the observed surface wind gusts. Derived images generated from the infrared dataset consisted of the following: (1) band 5-3 difference image; (2) band 4-5 difference image; (3) microburst risk image derived from the algorithm as defined in equation 1.  In the infrared derived images, contrast stretching and contouring of brightness was employed to highlight regions of relatively high microburst risk.  For each microburst event, product images were compared to radar reflectivity imagery and surface observations of convective wind gusts as provided by INL mesonet stations. Next Generation Radar (NEXRAD) base reflectivity imagery (levels II and III) from National Climatic Data Center (NCDC) was utilized to verify that observed wind gusts were associated with high-reflectivity downbursts and not associated with other types of convective wind phenomena (i.e. gust fronts). NEXRAD images were generated by the NCDC Java NEXRAD Viewer (Available online at http://www.ncdc.noaa.gov/oa/radar/jnx/index.html). Another application of the NEXRAD imagery was to infer microscale physical properties of downburst-producing convective storms. Particular radar reflectivity signatures, such as the rear-inflow notch (RIN)(Przybylinski 1995) and the spearhead echo (Fujita and Byers 1977), were effective indicators of the occurrence of downbursts.  In addition, radar temperature profiles over the INL were generated and archived as a means to validate quantitative information as portrayed in the derived product images.

      At this early stage in the algorithm assessment process, it is important to consider

covariance between the variables of interest: microburst risk (expressed as output brightness temperature) and surface downburst wind gust speed. A very effective means to assess the quantitative functional relationship between microburst index algorithm output and downburst wind gust strength at the surface is to calculate correlation between these variables. Thus, correlation between GOES imager microburst risk and observed surface wind gust velocities for the selected events were computed to assess the significance of these functional relationships. Statistical significance testing was conducted, in the manner described in Pryor and Ellrod (2004), to determine the confidence level of correlations between observed downburst wind gust magnitude and microburst risk values. Hence, the confidence level is intended to quantify the robustness of the correlation between microburst risk values and wind gust magnitude.

## 3. Case Studies

### 3.1 July 2007 Microbursts

The multispectral Geostationary Operational Environmental Satellite (GOES) imager microburst product has effectively indicated the potential for severe convective winds that were observed by Idaho National Laboratory (INL) mesonet stations in southeastern Idaho during July 2007. The microburst product indicated higher output brightness temperature (corresponding to higher microburst risk) about two hours prior to the occurrence of a severe convective wind gust (52 knots) on 31 July 2007 that was observed by Rover mesonet station. Lower brightness temperature was associated with a non-severe microburst (40 knots) that was observed by Specific Manufacturing Capability station in the INL on 25 July.

During the afternoon of 25 and 31 July 2007, convective storm activity developed over the Rocky Mountains of southeastern Idaho and propagated down slope to merge into clusters over the ESRP. Figure 2 displays GOES-11 imager derived microburst risk products at 2000 UTC 25 July 2007 and at 2030 UTC 31 July 2007, respectively. Note lower output brightness temperatures associated with the weaker downburst (40 knots, plotted in upper image) that occurred on 25 July as compared to the higher brightness temperatures associated with the severe microburst (52 knots, plotted in lower image) that occurred on 31 July. Also apparent in the images is convective storm activity, displayed as cellular or globular dark regions, developing over the surrounding Rocky Mountains. Forcing for this convective storm activity was most likely driven by solar heating of the mountain ridges in the presence of significant mid-level moisture that was advected from the central and eastern North Pacific Ocean. The convective storm activity propagated into the Snake River Plain where intense surface heating during the afternoon hours resulted in the development of a deep dry-adiabatic boundary layer. As discussed earlier, large BTD between bands 3 and 5 implies a large relative humidity gradient and well-mixed moisture profile between the mid-troposphere and the surface, a condition favorable for strong convective downdraft generation due to evaporational cooling of precipitation in the deep sub-cloud layer. Higher boundary layer moisture content observed during the afternoon of 25 July by Specific Manufacturing Capability mesonet station most likely resulted in decreased downdraft instability and the resulting weaker microburst.

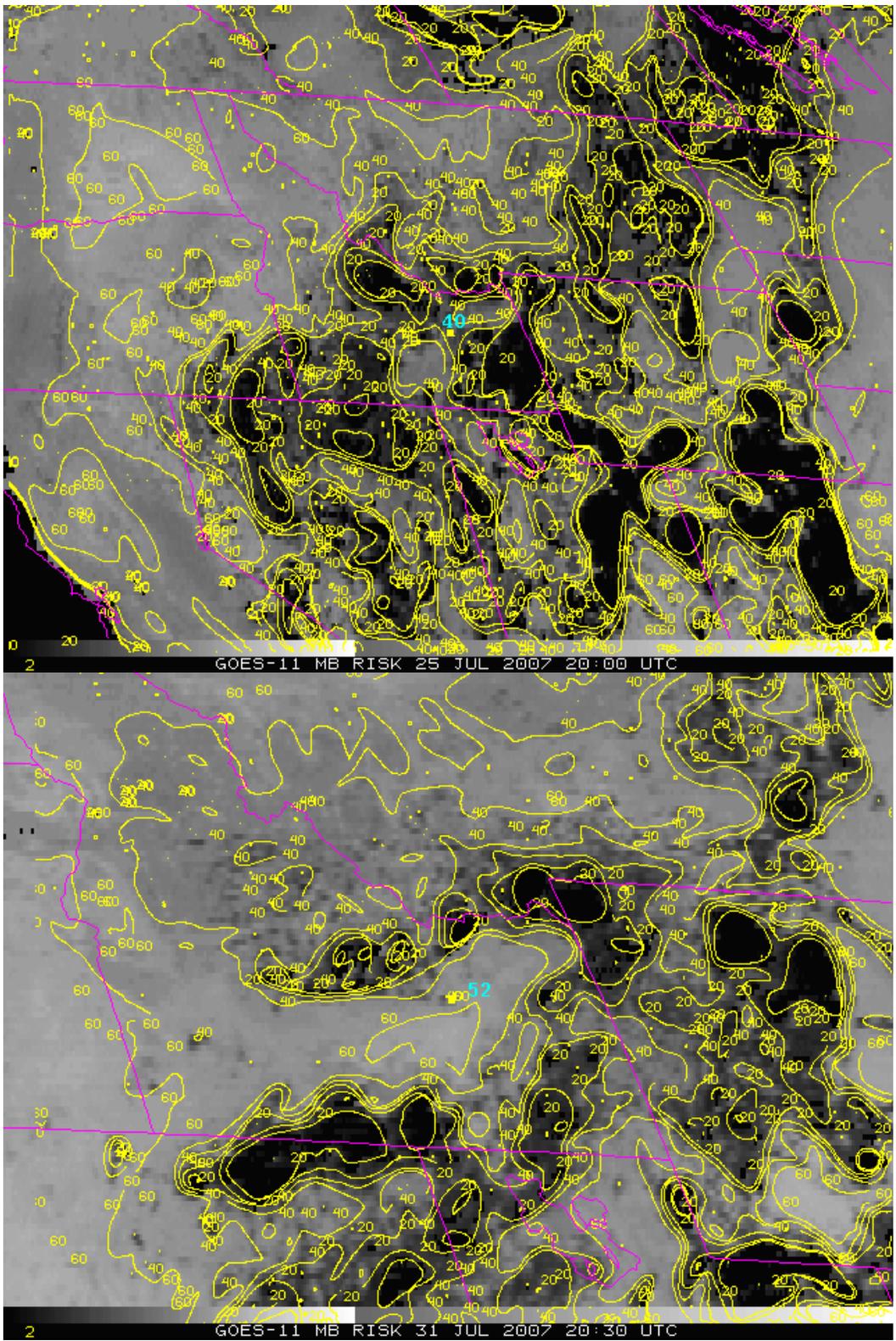

Figure 2. GOES-11 imager microburst risk products at 2000 UTC 25 July 2007 (top) and at 2030 UTC 31 July 2007 (bottom) with observed peak convective wind gusts plotted over the images. Output brightness temperature contours are overlying the image.

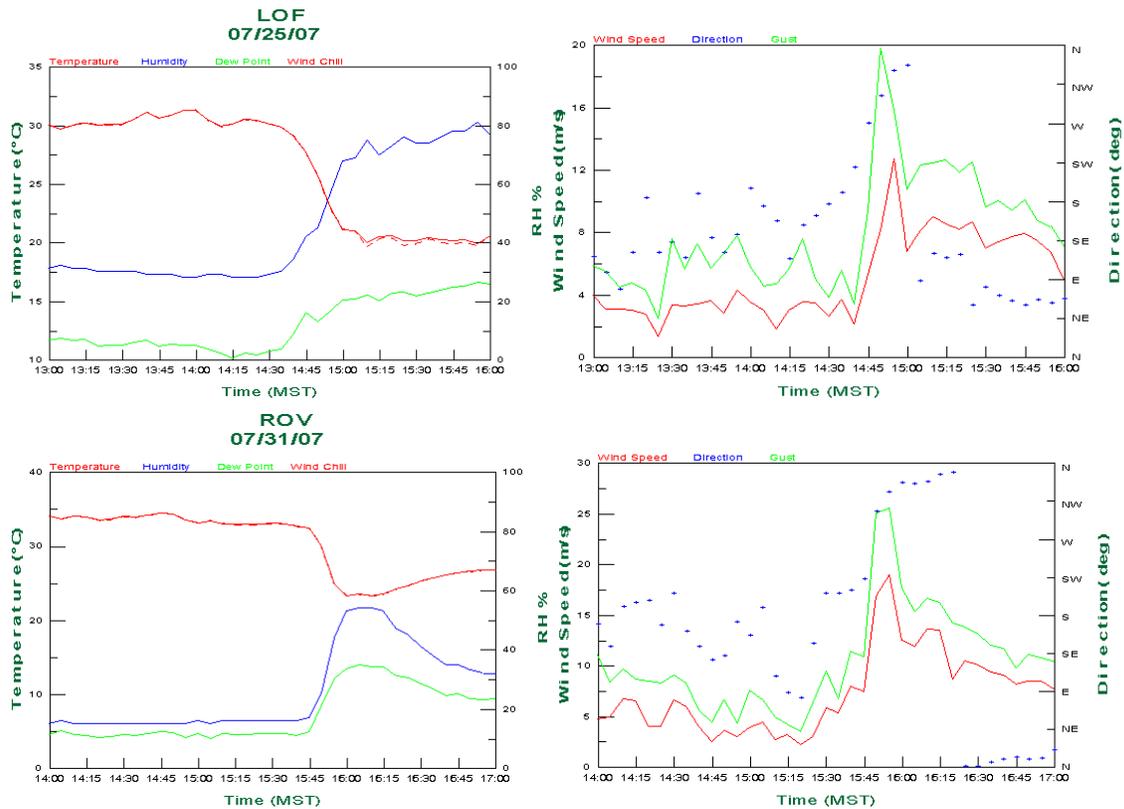

Figure 3. INL meteograms at Specific Manufacturing Capability mesonet station (LOF, top) and Rover mesonet station (ROV, bottom).

Surface observations as displayed in the meteograms in Figure 3 confirm the presence of favorable boundary layer conditions for microbursts during the afternoon hours of 25 and 31 July. Especially noteworthy is a large surface dewpoint depression near 29°C and relative humidity less than 20% during the two hours prior to downburst occurrence at Rover station. In addition, temperature lapse rates in the surface layer were determined to be superadiabatic (>10°C/km) based on comparisons between temperatures measured at 2 meters and 15 meters above ground level (not shown) about two hours prior to downburst occurrence. Superadiabatic lapse rates are often associated with strong insolation and resulting surface heating.

Downburst occurrence was apparent in the mesonet meteograms, characterized by sharp peaks in wind speed, and abrupt changes in wind direction, temperature and relative humidity. Radar imagery in Figure 4 from Pocatello NEXRAD (KSFX) near the time of occurrence of peak winds indicated that high reflectivities (>45 dBZ) were associated with the parent convective storms of downbursts. The combination of high radar reflectivity and large subcloud lapse rates strongly suggests that the microbursts observed during the afternoon of 25 and 31 July were hybrid microbursts. Also apparent were radar signatures typically associated with high-reflectivity microbursts such as the rear-inflow notch (RIN)(Przybylinski 1995) and the spearhead echo (Fujita and Byers 1977).

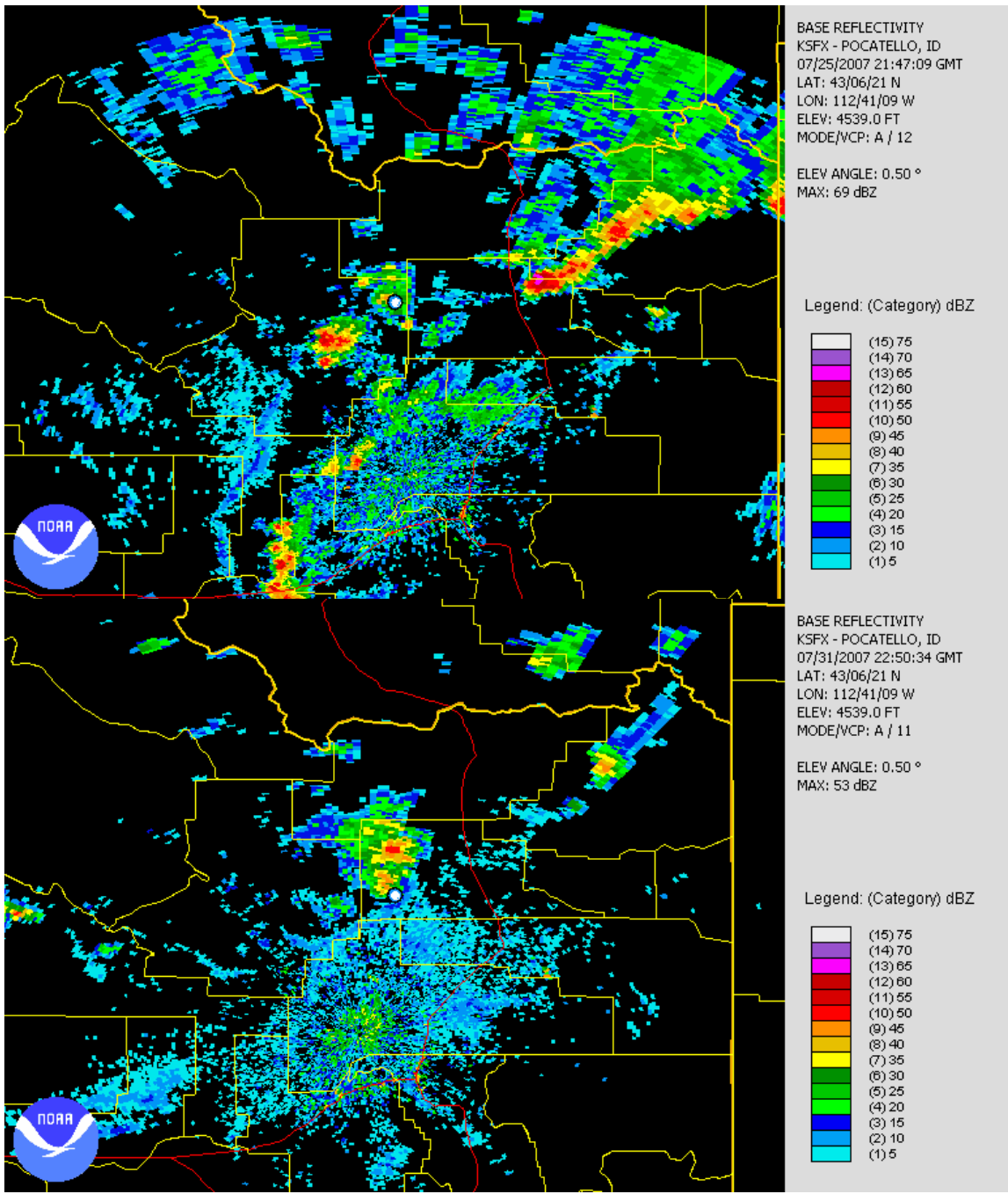

Figure 4. Radar reflectivity images from Pocatello NEXRAD at 2147 UTC 25 July 2007 (top) and 2250 UTC 31 July 2007 (bottom). Markers indicate the locations of Specific Manufacturing Capability (top) and Rover (bottom) mesonet stations, respectively.

## 3.2 August 2007 Microbursts

The most significant microburst event was recorded by San Dunes (SAN) and Rover (ROV) mesonet stations during the evening of 11 August 2007 with wind gusts of 60 knots. A cluster of convective storms developed over the ESRP and tracked northward over the INL during the late afternoon. Between 0030 and 0045 UTC 12 August, the cluster of convective storms produced severe downbursts at San Dunes and Rover mesonet stations. Similar to the pre-convective environments of the July microburst events, the convective boundary layer was deep and dry with superadiabatic temperature lapse rates near the surface. Exceptionally dry surface layer conditions were indicated in the San Dunes meteogram in Figure 5 with the dewpoint depression greater than 35°C and relative humidity less than 10% during the two hours prior to microburst occurrence. Figure 6, the GOES-11 microburst risk image at 2230 UTC 11 August 2007, reflected favorable microburst conditions with output brightness temperatures greater than 60 in close proximity to SAN and ROV stations.

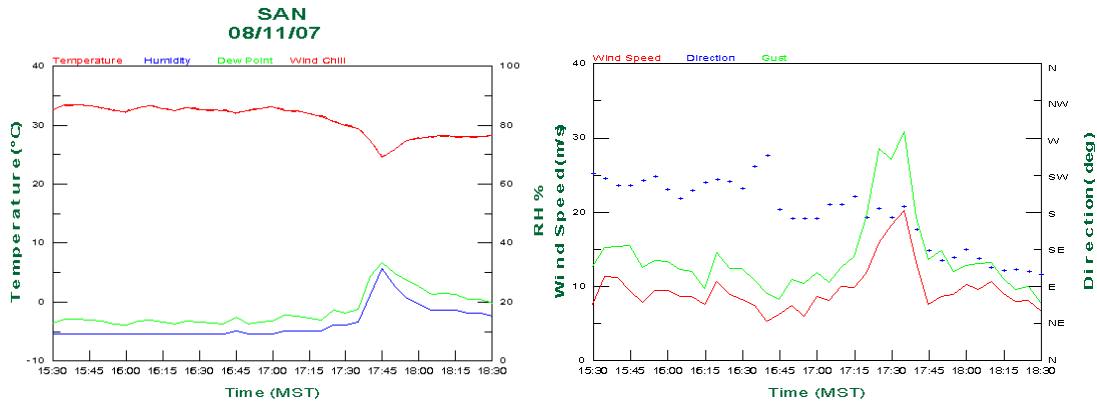

Figure 5. INL meteogram at San Dunes (SAN) mesonet station.

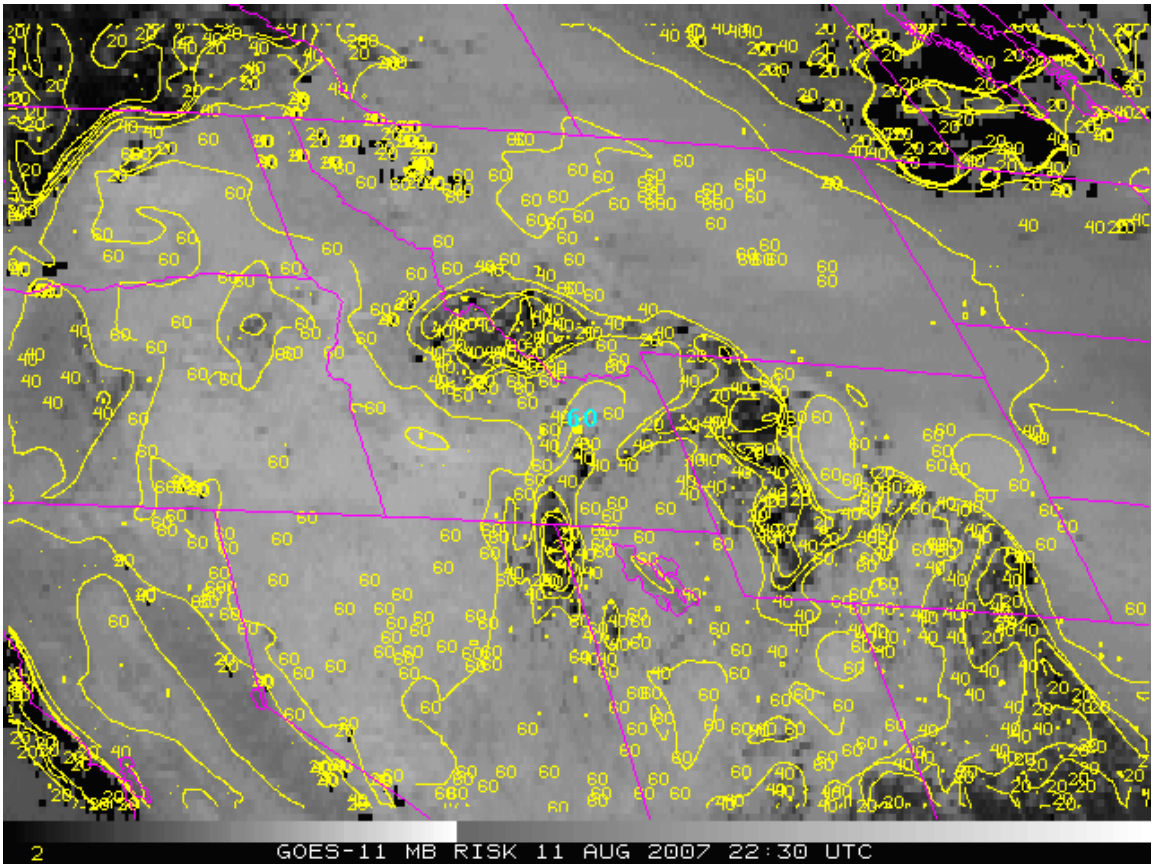

Figure 6. GOES-11 imager microburst risk product at 2230 UTC 11 August 2007 with observed peak convective wind gust at Sand Dunes station plotted over the image.

Also, similar to the July events, NEXRAD imagery in Figure 7 revealed reflectivities greater than 45 dBZ associated with the parent convective storms of the microbursts, again, suggesting that microbursts in this case could be classified as hybrid. As was typical for the 2007 convective season, mid-level moisture available for deep convective storms originated over the central North Pacific Ocean and was advected northeastward over southeastern Idaho. Interesting to note is the light amount of rainfall (.01 in.) measured by San Dunes mesonet station coincident with the microburst. Comparing high radar reflectivity with the measured rainfall amount may suggest that significant evaporation occurred as precipitation descended in the sub-cloud layer, thus, providing negative buoyant energy for downdraft acceleration.

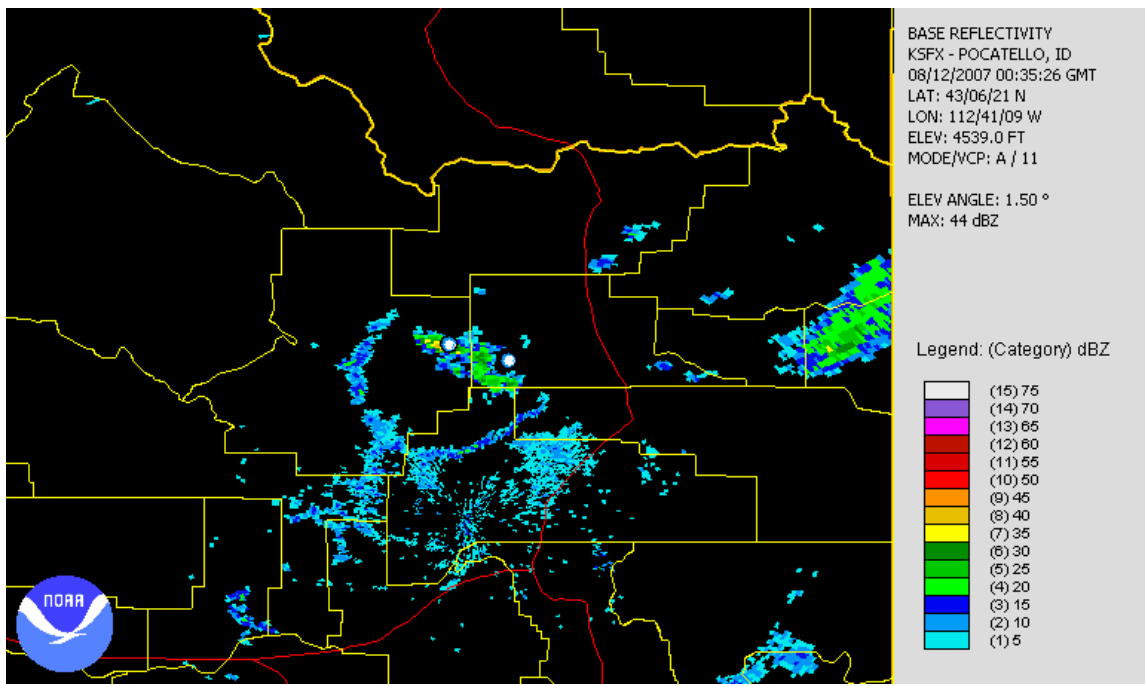

Figure 7.  Radar reflectivity image from Pocatello NEXRAD at 0035 UTC 12 August 2007.  Markers indicate the locations of San Dunes and Rover mesonet stations.

**4.  Statistical Analysis and Discussion**

   Analysis of covariance between the variables of interest, microburst risk (expressed as output brightness temperature) and surface downburst wind gust speed, provided favorable results for the imager microburst product.  A strong correlation (r=.76) between microburst risk values and wind gust speed was found for the dataset of ten microburst events during July and August 2007.  In addition, statistical significance testing revealed a high (98%) confidence level that the correlation did represent a physical relationship between risk values and downburst magnitude and was not an artifact of the sampling process.  A listing of the microburst events during July and August 2007 is displayed in Table 1.  Table 1 shows that microbursts in the INL mesonet domain occurred primarily during the afternoon.  This preference for afternoon microburst activity underscores the importance of solar heating of the boundary layer in the process of convective downdraft generation over southeastern Idaho.  The well-mixed moisture profile and relative humidity gradient that results from diurnal heating fostered a favorable environment for microbursts due to the evaporation of precipitation in the deep sub-cloud layer.  In addition, the majority of the observed microbursts, including the most intense recorded microburst (60 knots), were associated with high radar reflectivity (> 45 dBZ).  Significant mid-level moisture promoted precipitation loading as an initiating mechanism for downbursts over the INL domain.  The combination of precipitation loading and the presence of a relatively deep and dry convective boundary layer favored a microburst environment that was effectively captured by the GOES imager-derived microburst product.  Thus, derivation of an algorithm that incorporates

GOES-11 bands 3, 4, and 5 appears to be most effective in indicating a favorable thermodynamic environment for microbursts over southeastern Idaho as well as other regions in the intermountain western U.S.

**Correlation:**
MBR to measured wind:        0.7569
Reflectivity to measured wind:   -0.4226
No. of events:    10.    Mean Wind Speed (kt):   50.20
                         Mean MBR:              46.50

| Date | Time (UTC) | Measured Wind Speed kt | Location | GOES-11 MBR |
|---|---|---|---|---|
| 8-Jul-07 | 100 | 42 | GRI | 40 |
| 14-Jul-07 | 2310 | 58 | DUB | 50 |
| 25-Jul-07 | 2150 | 40 | LOF | 40 |
| 31-Jul-07 | 2250 | 52 | ROV | 55 |
| 3-Aug-07 | 2250 | 48 | LOF | 50 |
| 4-Aug-07 | 2010 | 54 | DUB | 45 |
| 5-Aug-07 | 1910 | 48 | ATO | 40 |
| 12-Aug-07 | 30 | 60 | SAN | 60 |
| 16-Aug-07 | 2340 | 54 | GRI | 45 |
| 31-Aug-07 | 2245 | 46 | MON | 40 |

Table 1. Microburst events as recorded by INL mesonet stations during July and August 2007.

## 5. Summary and Conclusions

A new multispectral GOES imager product has been developed to assess downburst potential over the western United States. This microburst risk product image incorporates GOES-11 bands 3, 4, and 5 to sample the warm-season pre-convective environment and derive moisture stratification characteristics of the boundary layer that would be relevant in the microburst potential assessment process. Case studies and statistical analysis for downburst events that occurred over southeastern Idaho during July and August 2007 demonstrated the effectiveness of the product with a strong correlation between risk values and microburst wind gust magnitude. This product provides a higher spatial (4 km) and temporal (30 minutes) resolution than is currently offered by the GOES sounder microburst products and thus, should provide useful information to supplement the sounder products in the convective storm nowcasting process.

## 6. References

Atkins, N.T., and R.M. Wakimoto, 1991: Wet microburst activity over the southeastern United States: Implications for forecasting. *Wea. Forecasting*, **6**, 470-482.

Clawson, K.L., N.R. Ricks, and G.E. Start, 1989: Climatography of the Idaho National Engineering Laboratory. Department of Energy, 169 pp.

Fujita, T.T., and H.R. Byers, 1977: Spearhead echo and downburst in the crash of an airliner. *Mon. Wea. Rev.*, **105**, 129–146.

Pryor, K.L., and G.P. Ellrod, 2004: WMSI - A new index for forecasting wet microburst severity. *National Weather Association Electronic Journal of Operational Meteorology*, 2004-EJ3.

Przybylinski, R.W., 1995: The bow echo. Observations, numerical simulations, and severe weather detection methods. *Wea. Forecasting*, **10**, 203-218.

Sorbjan, Z., 1989: Structure of the atmospheric boundary layer. Prentice Hall, 317pp.

Wakimoto, R.M., 1985: Forecasting dry microburst activity over the high plains. *Mon. Wea. Rev.*, **113**, 1131-1143.